\documentclass[prl,twocolumn]{revtex4}
\usepackage[T1]{fontenc}
\usepackage[latin9]{inputenc}
\usepackage{amssymb}

\makeatletter

\providecommand{\tabularnewline}{\\}

\makeatletter

\begin{document}
\def\order#1{{\mathcal{O}}\left( #1 \right)}

\preprint{Alberta Thy 06-08}

\title{Dipole Excitation of Dipositronium}

\author{Mariusz Puchalski and Andrzej Czarnecki}
\affiliation{Department of Physics, University of Alberta, Edmonton, Alberta,
Canada T6G 2G7}
\begin{abstract}
The energy interval between the ground and the P-wave excited states
of the recently discovered positronium molecule $\mathrm{Ps}_{2}$ is
evaluated, including the relativistic and the leading logarithmic
radiative corrections, $E_{\mathrm{P}}-E_{\mathrm{S}} =
0.181\,586\,7\left(8\right)\mathrm{a.u.}$.  The P-state, decaying
usually via annihilation, is found to decay into the ground state by
an electric dipole transition 19 percent of the time. Anticipated
observation of this transition will provide insight into this exotic
system.
\end{abstract}

\maketitle

Last year's discovery \cite{CassidyNat2007} of dipositronium
($\mathrm{Ps}_{2}$) was welcomed as a herald of a new kind of
chemistry. $\mathrm{Ps}_{2}$ is a bound state consisting of two
electrons and two positrons. Its stability against dissociation was
predicted in \cite{HylleraasPs2}, but its observation was very
challenging. $\mathrm{Ps}_{2}$ rapidly annihilates producing photons
similar to those from atomic Ps decays. Nevertheless, the evidence of
the $\mathrm{Ps}_{2}$ existence is now compelling.

In the experiment described in \cite{CassidyNat2007}, an intense pulse
of positrons is stopped in porous silica, forming Ps atoms, some of
which make their way into the voids of the pores.  Ps atoms have two
hyperfine states: a short-lived spin-singlet para-positronium (pPs)
with a lifetime in vacuum of $\tau_{\mathrm{pPs}}=0.125$ ns, and a
long-lived spin-triplet ortho-positronium (oPs), with
$\tau_{\mathrm{oPs}}=142$ ns. Interactions may shorten the oPs
lifetime through two mechanisms: spin exchange quenching (SEQ), in
which spins flip converting oPs into the rapidly-decaying pPs; and
formation of molecules $\mathrm{Ps}_{2}$. In the latter case, as we
will discuss below, the probability of each of the electron-positron
pairs to be a spin-singlet is one quarter, and the size of the
molecule is similar to that of an atom. Thus $\mathrm{Ps}_{2}$
is short-lived, with the lifetime of about
$2\tau_{\mathrm{pPs}}$.

But even if the rapid disappearance of oPs is observed, how can one
tell whether this is due to molecule formation rather than SEQ? The
key is that the molecule formation needs another body, such as the
pore surface, to absorb the released binding energy.  As the
temperature of silica is increased, the fraction of Ps atoms on the
surface decreases, fewer molecules should be formed, and more oPs
survive.  Exactly such an effect is observed
\cite{CassidyNat2007}. More recently, evidence of the
$\mathrm{Ps}_{2}$ formation on a metal surface has also been found
\cite{cassidy:062511}.

As well as proving the existence of the first known system containing
more than one positron, this discovery is viewed as an important step
towards studies of even more exotic phenomena: larger polyelectrons,
Bose condensation of positronia, and eventually a $\gamma$-ray
laser based on stimulated annihilation.

Before those exciting possibilities are explored, a more detailed
study of the newly-discovered $\mathrm{Ps}_{2}$ is warranted. Like
the relatively better-known positronium ion $\mathrm{Ps}^{-}$ \cite{fleischer2006,mills1981},
the molecule $\mathrm{Ps}_{2}$ is weakly bound \cite{LeeVashishtaKalia83,Ho83}.
However, whereas the ion has no excited states stable against dissociation,
the molecule has an interesting spectrum of three: two excited S states
and one P \cite{VargaSecondState,Kinghorn1993}. The latter has a
sizable branching ratio for an electric dipole transition to the ground
state, accessible to middle-ultraviolet laser spectroscopy \cite{UsukuraSignature}.
This is a great asset since polyelectrons generally decay through
annihilation into photons, too much like atomic pPs to reveal the
structure of the decaying system. Anticipated observation and precise
measurement of this line \cite{MillsPrivate} will confirm the presence
of the molecules and test our understanding of their nature.

Measurements of the P-state excitation energy are expected to have a
precision of 10 parts per million (ppm) \cite{MillsPrivate}, sensitive
to relativistic corrections. The relativistic effects have been found
recently for the ground state \cite{bubin2007} but the P-level energy
is only known in the non-relativistic approximation
\cite{VargaSecondState}.

The challenge of precise theoretical studies of $\mathrm{Ps}_{2}$ is
that it is a relatively complicated four-body system whose components
have equal masses. Unlike in molecules built of atoms with hadronic
nuclei, no part of $\mathrm{Ps}_{2}$ can be treated as slowly-varying
and the Born-Oppenheimer approximation cannot be
applied. Nevertheless, its ground state has already been thoroughly
studied with variational and quantum Monte Carlo methods (see
\cite{emami-razavi2008} for a recent review of earlier work). The
P-state is more challenging.  Because its wave function has a node,
the variational procedure converges slower. Larger expressions for the
matrix elements and additional integrals exacerbate the difficulties
\cite{HarrisGauss,PachuckiGauss}.

To overcome these obstacles, we combined the variational method in
a Gaussian basis with algorithms for decomposing the Hamiltonian matrix
and for optimizing the wave function. In addition, we sped up the
convergence by transforming the operators representing corrections
to the energy, using a method proposed by Drachman \cite{Drachman81}.
As a result we not only match or exceed the accuracy of the best existing
evaluation of the relativistic corrections to the ground state, but
also extend those results to obtain the leading next-order corrections
(QED) and, more important, determine analogous effects for the P-state.
We find the energy interval 
\begin{equation}
\Delta E\equiv E_{\mathrm{P}}-E_{\mathrm{S}} = 
0.181\,586\,7\left(8\right)\mathrm{a.u.},
\label{eq:interval}
\end{equation}
or $4.941\,23\left(2\right)$ eV, corresponding to the wavelength
$\lambda=250.9179\left(11\right)$ nm. The branching ratio for the
dipole transition is also determined,
\begin{equation}
\mathrm{BR}\left(\mathrm{P\to\mathrm{S}}\right)
\equiv
\frac{\Gamma_{\mathrm{dip}}\left(\mathrm{P\to\mathrm{S}}\right)}{\Gamma_{\mathrm{annih}}
\left(\mathrm{P}\right)+\Gamma_{\mathrm{dip}}\left(\mathrm{P\to\mathrm{S}}\right)}
=0.191\left(2\right).
\label{eq:branching}
\end{equation}

The dipositronium's Hamiltonian is 
\begin{equation}
H=H_{\mathrm{C}}+\alpha^{2}H_{\mathrm{rel}}+\order{\alpha^{3}\ln\alpha}.\label{eq:hamiltonian}\end{equation}
Its leading term describes the non-relativistic Coulomb system,\begin{equation}
H_{\mathrm{C}}=\sum_{a=1}^{4}\frac{\vec{p}_{a}^{2}}{2}+\sum_{a<b}\frac{z_{ab}}{r_{ab}},\label{eq:Coulomb}\end{equation}
 where $\vec{p}_{a}$ and $r_{ab}\equiv\left|\vec{r}_{a}-\vec{r}_{b}\right|$
are momenta and relative distances of positrons ($a,b=1,\, 2$), and electrons
(3, 4). $z_{ab}$ equals $+1$ for a like-charged pair $a,\, b$
and $-1$ for opposite charges. As units of length, momentum, and
energy, we use $1/\alpha m_{e}$, $\alpha m_{e}$, and $\alpha^{2}m_{e}$,
and set $c=\hbar=1$.

Wave functions and the lowest-order ({}``nonrelativistic'') values
of energy levels are determined by the Coulomb Hamiltonian (\ref{eq:Coulomb}).
Higher-order corrections $\order{\alpha^{2},\alpha^{3}\ln\alpha}$
to energies are computed as first-order perturbations with those wave
functions.

The Hamiltonian (\ref{eq:hamiltonian}) has a rich symmetry 
\cite{Kinghorn1993,schrader2004}
that is reflected in the wave functions. In addition to the symmetry
with respect to permuting coordinates of electrons, as well as those
of positrons, there is also the charge conjugation symmetry. If the
system is described by relative coordinates $r_{ab}$ only, the latter
is equivalent to the spatial inversion, and described in a given state
by its parity $\pi$. Thus the wave functions of the ground state
with $L^{\pi}=0^{+}$ and of the P-state $1^{-}$ are 
$\psi_{i}=\mathcal{A}\left[\chi\phi_{i}\right]$,
 where $\chi=\frac{1}{2}\left(\uparrow_{1}\downarrow_{2}-\downarrow_{1}\uparrow_{2}\right)\left(\uparrow_{3}\downarrow_{4}-\downarrow_{3}\uparrow_{4}\right)$
is constructed using electron and positron spinors and
the antisymmetrizer  is built out of operators
permuting pairs of particles, $\mathcal{A}=\frac{1}{\sqrt{8}}\left(1+\pi P_{13}P_{24}\right)\left(1-P_{12}\right)\left(1-P_{34}\right)$.
The spatial wave functions $\phi_{i}$ are expressed in a Gaussian basis,
\begin{eqnarray}
\phi_{\mathrm{S}} & = & \sum_{i=1}^{N}c_{i}^{\mathrm{S}}\exp\left[-\sum_{a<b}w_{ab}^{iS}r_{ab}^{2}\right],\nonumber \\
\phi_{\mathrm{P}} & = & \vec{r}_{1}\sum_{i=1}^{N}c_{i}^{\mathrm{P}}\exp\left[-\sum_{a<b}w_{ab}^{iP}r_{ab}^{2}\right].\label{eq:basis}\end{eqnarray}
 Here $\vec{r}_{i}$ denotes a particle coordinate with respect to
the center of mass. Since $\vec{r}_{1}+\vec{r}_{2}+\vec{r}_{3}+\vec{r}_{4}=0$,
in terms of the relative coordinates we have $\vec{r}_{1}=\frac{1}{4}\left(\vec{r}_{12}+\vec{r}_{13}+\vec{r}_{14}\right)$.

The six parameters $w_{ab}^{i\,\mathrm{S,P}}$ in (\ref{eq:basis}) are
determined, for each of the $N$ elements of the basis, in an extensive
optimization process.  QR decomposition \cite{NumRecipes} and inverse
iteration are used to determine energy eigenvalues of $H_C$.  In the
$i^{\mathrm{th}}$ step of minimizing the energy, the six parameters of
the $i^{\mathrm{th}}$ basis element are optimized with Powell's method
without gradient \cite{NumRecipes}.  The optimization steps are cycled
through the basis elements until convergence is reached.  This
procedure yields nonrelativistic energies accurate to better than one
part per billion, shown in the first line of Table \ref{tab:levels}.
The ground state agrees with \cite{bubin2007}, although our error bar
is slightly larger. For the P-state, we improve on the previous best
result \cite{VargaSecondState}.

 To test the numerical procedure, we used the ground-state of lithium,
presently the best known four-body system. Its ground-state energy was
evaluated using Hylleraas coordinates
\cite{PuchalskiLit,PachuckiRem,pachucki:032514} with a precision of
about $10^{-12}$. It reliably calibrates both the absolute value and
its uncertainty since, thanks to the richer symmetry of Ps$_{2}$, the
Gaussian method converges better for it than for Li.

Assured of the quality of the obtained wave function, we proceed to
the relativistic corrections. For the two states of interest, the
following parts of the Breit-Pauli Hamiltonian \cite{Berestetsky:1982aq}
contribute, \begin{eqnarray}
H_{\mathrm{rel}} & = & H_{\mathrm{MV}}+H_{\mathrm{D}}+H_{\mathrm{OO}}+H_{\mathrm{SS}}+H_{\mathrm{A}},\label{eq:breit}\\
H_{\mathrm{MV}} & = & -\frac{1}{8}\sum_{a}\vec{p}_{a}^{4},\\
H_{\mathrm{D}} & = & -\pi\sum_{a<b}z_{ab}\;\delta^{3}\left(r_{ab}\right),\\
H_{\mathrm{OO}} & = & -\frac{1}{2}\sum_{a<b}z_{ab}\; p_{a}^{i}\left(\frac{\delta^{ij}}{r_{ab}}+\frac{r_{ab}^{i}r_{ab}^{j}}{r_{ab}^{3}}\right)p_{b}^{j},\\
H_{\mathrm{SS}} & = & -\frac{2\pi}{3}\sum_{a<b}z_{ab}\;\vec{\sigma}_{a}\cdot\vec{\sigma}_{b}\;\delta^{3}\left(r_{ab}\right),\\
H_{\mathrm{A}} & = & \frac{\pi}{2}\sum_{a<b,ab\neq12,34}\left(3+\vec{\sigma}_{a}\cdot\vec{\sigma}_{b}\right)\delta^{3}\left(r_{ab}\right).\end{eqnarray}
 Given the spin configuration of Ps$_{2}$, $\vec{\sigma}_{a}\cdot\vec{\sigma}_{b}$
can be replaced by zero for an $e^{+}e^{-}$ pair, and by $-3$ for
like-charged pairs \cite{bubin2007}. Thus, only four operators remain:
$p^{4}$, delta-functions for $e^{+}e^{-}$ and $e^{-}e^{-}$, and
$H_{\mathrm{OO}}$.

A disadvantage of the Gauss basis is its incorrect behavior at short
inter-particle distances: it does not reproduce the cusps of the wave
function. This slows down the convergence of matrix elements of the
delta-function and the kinetic energy ({}``mass-velocity'') correction
$p^{4}$. For example, in lithium with a basis size of 2000, the error
is a few units in $10^{4}$ \cite{pachucki:cusp}. Even more dangerously,
the convergence is so slow that a misleading limit may be deduced.
To overcome this difficulty, the operators can be transformed into
an equivalent form, whose behavior is less sensitive to the shortest
distances. For the delta-function, a prescription was found by Drachman
\cite{Drachman81}. Neglecting boundary terms, 
$ 4\pi\delta^{3}\left(r_{ab}\right)\phi_{1}\phi_{2} \to
\frac{2}{r_{ab}}\left(E-V\right)\phi_{1}\phi_{2}
+\sum_{c}\frac{\left(\nabla_{c}^{i}\phi_{1}\right)
\left(\nabla_{c}^{i}\phi_{2}\right)}{r_{ab}}$,
where $V=\sum_{a<b}\frac{z_{ab}}{r_{ab}}$. For the kinetic energy we
use $\sum_{a}\phi_{1} p_{a}^{4}\phi_{2}
\to 4\left(E-V\right)^{2}\phi_{1}\phi_{2}
-2\sum_{a<b}\left(\nabla_{a}^{2}\phi_{1}\right)\left(\nabla_{b}^{2}\phi_{2}\right)$.

Numerical results for the four basic relativistic operators are shown
in Table \ref{tab:fourBasicOperators}. For the ground state we find
agreement with \cite{bubin2007} to within 1 ppm for $H_{\mathrm{OO}}$,
and three or four digits for the remaining operators.  Since these are
the operators that we regularized, we believe our results to be more
accurate, despite the much smaller size of the basis used (the
difference is unimportant since unknown higher-order effects are
likely larger).  Again, lithium was used as a test.
The kinetic correction converges slowest, as usual in
many-body calculations
\cite{PuchalskiLit}.

Table \ref{tab:levels} shows corrections to the energy levels.
The orbit-orbit and the Darwin terms largely cancel. The result is
dominated by the virtual annihilation and, to a smaller degree, by
the kinetic term. The annihilation is repulsive and decreases the
binding. It is about twice as effective in the ground state as in
the P-state so that overall the relativistic effects decrease
the S-P interval.

Beyond relativistic corrections, the next largest effect is of the
relative size $\order{\alpha^{3}\ln\alpha}$, analogous to the Lamb
shift in hydrogen. Its physics is richer in positronium \cite{Fulton}
because two-photon {}``recoil'' interactions between two light
constituents contribute at the same order as the self-interaction
corrections.  In hydrogen the latter dominate, the recoil effects
being suppressed by the electron-to-proton mass ratio. In Ps$_{2}$,
one should in principle consider interactions among all pairs of
particles. However, contributions of like-charged particles are
suppressed by two orders of magnitude, as can be seen by comparing the
last two columns of Table \ref{tab:fourBasicOperators}. Thus the
coefficient of $\alpha^{3}\ln\alpha$ is given with excellent accuracy
by expectation values of $ \alpha^{3}\ln{\alpha}\;
H_{\mathrm{log}}\equiv-24\alpha^{3}\ln{\alpha}\;\delta^{3}\left(r_{13}\right)$
\cite{Pineda:1998kn}, shown in Table \ref{tab:levels}.  We take halves
of their values to estimate the error for each level, and add those to
obtain the error estimate of the final result, Eq.
(\ref{eq:interval}).

The precise value of the $e^{+}e^{-}$ overlap given in Table
\ref{tab:fourBasicOperators} provides a new prediction of the
annihilation rate of Ps$_{2}$ in both states, 
\begin{eqnarray}
\Gamma_{\mathrm{annih}}({\mathrm{S}}) & = & 4\pi\alpha^{3}\left\langle
\delta^{3}\left(r_{13}\right)\right\rangle (1+\mathrm{RC})\nonumber \\
& = & 1/(0.224\,55(6)\;\mathrm{ns}),\label{annihS}\\
\Gamma_{\mathrm{annih}}({\mathrm{P}}) & = &
1/(0.442\,77(11)\;\mathrm{ns}),\label{annihP}
\end{eqnarray}
 where the RC denotes the radiative corrections, 
\[
\mathrm{RC}=\alpha\left(\frac{19\pi}{12}-\frac{17}{\pi}\right)+2\alpha^{2}\ln\frac{1}{\alpha}
+\order{\alpha^{2}}.\]
The error, due to unknown $\order{\alpha^{2}}$ effects, is estimated
as half of the logarithmic correction. The annihilation in state S was
previously calculated in \cite{bailey2005}. Our result differs
slightly, primarily because of the error in the $\order{\alpha}$
correction in that study. The result (\ref{annihS}) confirms the
expectation that the lifetime of Ps$_{2}$ in its ground state is about
half of that in pPs. In the P-state, one of the particles overlaps
only negligibly with the others, slowing down the annihilation by
another factor of two.

For the experimental search of the dipole transition, it is
interesting to know how competitive it is relative to the
annihilation. The rate of the dipole transition is \begin{eqnarray}
\Gamma_{\mathrm{dip}}({\mathrm{P}\to\mathrm{S}}) & = &
\frac{4}{3}\alpha^{3}\left(E_{\mathrm{P}}-E_{\mathrm{S}}\right)^{3}\left|\left\langle
\mathrm{S}\left|\vec{d}\right|\mathrm{P}\right\rangle
\right|^{2}\nonumber \\ & = &
1/\left(1.873\;\mathrm{ns}\right),\end{eqnarray} where
$\vec{d}\equiv{\vec{r}}_{1}+{\vec{r}}_{2}-{\vec{r}}_{3}-{\vec{r}}_{4}$.
The dipole matrix element is determined as a Gaussian integral,
$\left|\left\langle
\mathrm{S}\left|\vec{d}\right|\mathrm{P}\right\rangle
\right|=2.040\,942\,265(16)$, and leads to the final result for the
branching ratio, Eq.~(\ref{eq:branching}).  Corrections to this
prediction, conservatively estimated as less than one percent, arise
from the three-body decay into a photon and two Ps atoms, and from
relativistic effects $\order{\alpha^{2}}$ \cite{DrakeRelCor}. The
present value exceeds the previous evaluation \cite{UsukuraSignature}
by about ten percent, a welcome improvement for the experimental
search of this transition. The larger value may slightly facilitate
the use of the P-S transition intensity to determine the number of the
Ps$_{2}$ molecules produced.

It is noteworthy that the rate of the dipole transition is similar to
twice that in a positronium atom. This confirms the approximate
picture of the excited P-state as resembling two weakly interacting
positronium atoms, one in its ground state and the other in the
P-state \cite{VargaSecondState}. In a molecule consisting of two
weakly interacting atoms, the P-S dipole matrix element is, because of
coherence, $\sqrt{2}$ times larger than in an isolated atom. This
approximation predicts
$|\langle\vec{d}\rangle|=\frac{512}{243}=2.1$. Similarly, in molecular
hydrogen one finds \cite{Wolniewicz2003}
$|\langle\vec{d}\rangle|=\frac{256}{243}=1.05$, in the limit of weak
interaction between the two atoms.

Fortunately, the 2P-1S energy interval does differ sufficiently between
atomic and molecular positronia for its measurement to unambiguously
confirm the existence of Ps$_{2}$. The main difference arises already
in the non-relativistic energy values. In the molecule, one Ps atom
may be interpreted as a dielectric medium that weakens the electric
field in the other one, thus decreasing all energy intervals. Relativistic
effects, primarily the annihilation, slightly add to that decrease.
The dipole matrix element is also decreased below the asymptotic value
of $\sqrt{2}$ times that in a free atom, weakening the transition
rate below half of the atomic rate.

On the technical side, this study reveals the somewhat unexpected
potential of the correlated Gaussian basis. The fast optimization
method described here leads to comparable or better results than previously
published, even with a much smaller basis. A drawback of Gaussians
is their incorrect asymptotic behavior, both at short and at long
distances. This is compensated by the availability of an analytical
form of all required matrix elements and by the good numerical behavior
of the integrals. Double precision sufficed for the variational parameters.

The Gaussian basis is especially suitable for the positronium molecule
since it tracks all inter-particle distances. 
The high symmetry of Ps$_{2}$ improves the convergence of the variational
procedure. Parallelizing the code would certainly lead to improvements,
and will likely be necessary for the determination of further QED
corrections. For the present and foreseeable measurement goals, the
theoretical description of the dipole transition energy and its probability
presented here is sufficient. Its experimental test will complement
the newest chapter in chemistry with one in spectroscopy.

\bigskip{}

\begin{acknowledgments}
We are indebted to Allen Mills, Jr., 
Krzysztof Pachucki, Alexander Penin, and Mikhail Voloshin
for very helpful discussions. We thank Alexander Brown and Lutos{\l{}}aw
Wolniewicz for advice on the molecular hydrogen literature, and Paul
McGrath for improving the manuscript. This research was supported
by Science and Engineering Research Canada. 
\end{acknowledgments}

\begin{table}
\begin{ruledtabular} \begin{tabular}{lll}
Source  & Ground state  & P state\tabularnewline
\hline 
$H_{\mathrm{C}}$  & $-0.516\,003\,790\,415\left(88\right)$  & $-0.334\,408\,317\,34(81)$\tabularnewline
$\alpha^{2}H_{\mathrm{MV}}$  & $-0.000\,009\,152$  & $-0.000\,004\,780\left(1\right)$\tabularnewline
$\alpha^{2}H_{\mathrm{OO}}$  & $-0.000\,013\,470$  & $-0.000\,007\,736$\tabularnewline
$\alpha^{2}H_{\mathrm{D}}$  & $\hphantom{-}0.000\,014\,592$  & $\hphantom{-}0.000\,007\,458$\tabularnewline
$\alpha^{2}H_{\mathrm{SS}}$  & $\hphantom{-}0.000\,000\,419$  & $\hphantom{-}0.000\,000\,097$\tabularnewline
$\alpha^{2}H_{\mathrm{A}}$  & $\hphantom{-}0.000\,022\,202$  & $\hphantom{-}0.000\,011\,259$\tabularnewline
\hline 
$\alpha^{2}H_{\mathrm{rel}}$  & $\hphantom{-}0.000\,014\,591$  & $\hphantom{-}0.000\,006\,298(1)$\tabularnewline
$\alpha^{3}\ln{\alpha}\; H_{\mathrm{log}}$  & $\hphantom{-}0.000\,001\,01(50)$  & $\hphantom{-}0.000\,000\,51(25)$\tabularnewline
\hline 
Total  & $-0.515\,988\,2(5)$  & $-0.334\,401\,5(3)$\tabularnewline
\hline 
\cite{bubin2007}  & $-0.515\,989\,199\,656$  & \tabularnewline
\end{tabular}\end{ruledtabular} 

\caption{Corrections to the energy levels of Ps$_{2}$.}
\label{tab:levels}
\end{table}

\begin{table*}
\begin{ruledtabular} \begin{tabular}{lllll}
Basis size  & $\left\langle \sum_{a}\vec{p}_{a}^{4}\right\rangle $  & $\left\langle \sum z_{ab}p_{a}^{i}\left(\frac{\delta^{ij}}{r_{ab}}+\frac{r_{ab}^{i}r_{ab}^{j}}{r_{ab}^{3}}\right)p_{b}^{j}\right\rangle $  & $10^{2}\left\langle \delta^{3}\left(r_{13}\right)\right\rangle $  & $10^{4}\left\langle \delta^{3}\left(r_{12}\right)\right\rangle $ \tabularnewline
\hline 
\multicolumn{5}{c}{Ground state}\tabularnewline
\hline 
2200  & $1.374\,923(45)$  & $0.505\,892\,400(27)$  & $2.211\,851\,17(14)$  & $6.256\,827\,3(42)$\tabularnewline
1600 \cite{UsukuraSignature}  &  &  & $2.211\,51$  & $6.259$\tabularnewline
6000 \cite{bubin2007}  & $1.374\,696\,3$  & $0.505\,892\,40$  & $2.211\,775\,9$  & $6.257\,950\,5$\tabularnewline
\hline 
\multicolumn{5}{c}{State P}\tabularnewline
\hline 
2200  & $0.718\,150(86)$  & $0.290\,557\,920(46)$  & $1.121\,723\,38(31)$  & $1.453\,512\,7(82)$\tabularnewline
1600 \cite{UsukuraSignature}  &  &  & $1.120\,91$  & $1.459\,1$\tabularnewline
\end{tabular}\end{ruledtabular}

\caption{Expectation values of the basic relativistic operators, compared with
previous studies, where available.}
\label{tab:fourBasicOperators}
\end{table*}


\end{document}